\documentclass[a4paper,nobibnotes, longbibliography, reprint,aip,rsi]{revtex4-1}

\usepackage[utf8]{inputenc}
\usepackage[T1]{fontenc}

\usepackage{amsmath}
\usepackage{graphicx}

\usepackage{gnuplottex}
\usepackage{tikz}
\usetikzlibrary{arrows, shapes, calc}

\usepackage{url}
\usepackage{hyperref}

\usepackage{booktabs}   

\usepackage{color}

\usepackage{adjustbox}  
\usepackage{import} 

\newcommand{\diff}{\;\mathrm{d}}

\usepackage{natbib}

\begin{document}

\title{Analyzing X-Ray tomographies of granular packings}
\date{\today}
\author{Simon \surname{Weis}}
\affiliation{Institut für Theoretische Physik I, Friedrich-Alexander-Universität, 91058 Erlangen, Germany}
\author{Matthias \surname{Schröter}}
\affiliation{Institute for Multiscale Simulation, Friedrich-Alexander-Universität, 91052 Erlangen, Germany.}
\begin{abstract}
Starting from three-dimensional volume data of a granular packing, as e.g.~obtained by X-ray Computed Tomography, 
we discuss methods to first detect the individual particles in the sample and then analyze their properties. 
This analysis includes the pair correlation function, the volume and shape of the Voronoi cells and the number 
and type of contacts formed between individual particles. We mainly focus on packings of monodisperse spheres, 
but we will also comment on other monoschematic particles such as ellipsoids and tetrahedra. 
This paper is accompanied by a package of free software containing all programs (including source code) and an 
example three-dimensional dataset which allows the reader to reproduce and modify all examples given.
\end{abstract}

\maketitle

\section{Introduction}

As most granular materials are opaque for light in the visible spectrum, standard CCD cameras will only obtain 
information from the surface of a granular sample. But this surface behavior is known to differ 
substantially from the bulk properties.\cite{zhang:06,orpe:07,jerkins:08,desmond:09}  
In contrast, X-rays will penetrate most granular samples with an intensity exponentially decaying with the length of the penetrated material, the so called Lambert-Beer law.\cite{lambert:1760,beer:1852}
Images obtained by placing a sample between and X-ray source and a camera with a scintillator are called 
radiograms. These have e.g.~been used  to study density variations in flowing sand\cite{michalowski:84,baxter:89,kabla:05,kabla:09}
or motion of larger objects inside a granular sample.\cite{royer:05,maladen:09,cao:14,homan:15}

If these radiograms are collected while rotating the sample over a sufficiently large angle
\footnote{180$^\circ$  for a parallel X-ray beam, as e.g.~delivered by a synchrotron. 
360$^\circ$ if the beam expands from a point source, as in a normal X-ray tube.},
a number of mathematical algorithms can be used to reconstruct the sample in three-dimensions,\cite{buzug:08}
a process called Computed Tomography.  The volume data obtained this way consist of individual voxels, a linguistic blend of the words volume and pixel. 
Each voxel represents the x-ray absorption coefficient within a little cube at the corresponding location in the 
sample.\footnote{The term voxel is used in two ways: to refer to the cuboidal image element itself and as a 
unit of length equivalent to one edge of this cube.}

Driven initially by medical research and then also material science, X-ray tomography has become a turnkey
solution for three-dimensional imaging. In fact, it is now even possible to assemble your own custom-made X-ray 
tomography system.\cite{athanassiadis:14} 
However, going beyond impressive images created with the
3D visualization software provided by the manufacturers does require serious image processing. 
The main objective of this article is to facilitate this step by providing a novice-friendly
description how to identify the individual particles in a granular sample and then to compute 
both global statistical quantities and local quantitative measures.

The parameters discussed in this paper are however by far not 
the only properties that can be accessed from  volume data. In the last 
decade X-ray tomography has been used to
analyze the local structure of packings,\cite{seidler:00,richard:03,aste:05,aste:05a,aste:07,tsukahara:08,al-raoush:10,gillman:13,xia:14,heim:16,baranau:16}  understand their mechanical stability,\cite{scheel:08,brown:12,cao:13} and the forces at interparticle contacts.\cite{delaney:10,saadatfar:12,hurley:16}
More dynamical properties studied include the formation of shear bands,\cite{hall:10,ando:12,ando:13,tordesillas:13,alikarami:15,hemmerle:16} the flow and compaction of rods,\cite{fu:12,borzsonyi:12,borzsonyi:12a,wegner:12,yadav:13,wegner:14,wortel:15,borzsonyi:16} segregation,\cite{finger:15,schella:17} order-disorder transitions,\cite{francois:13,hanifpour:14,hanifpour:15} granular media as a model system for glassy behavior,\cite{li:14,xia:15} and  the crushing of individual sand grains.\cite{alikarami:15,zhao:15} 

The remaining article is structured as follows: 
Section \ref{sec:visual} describes how to visualize and inspect the raw volume data. 
Section \ref{sec:find_particles} provides the foundation for all subsequent analysis steps
by describing how to identify the coordinates (and possibly orientations) of all  the individual particles
in the volume data. These coordinates will then be used to compute 
pair correlation functions (section \ref{sec:pair_correlation}), contact numbers (section \ref{sec:contacts}), 
and Voronoi volumes (section \ref{sec:voronoi}).

This paper is meant to be self-contained. It allows the reader to test all methods and reproduce all results
by providing a) a demo volume data set of a sphere packing, which can be downloaded from \cite{demo_url}, 
and b) the complete software package (as supplementary material and on github). The software
should compile on any standard 64 Bit Linux PC, for convenience we provide also
static linked binaries for all analysis step.
The technical details how to obtain and run the analysis programs can be found in appendix
\ref{sec:app_analyzing}.

While most of the paper pertains to the 
analysis of a sphere packings, we will also comment on packings of ellipsoids and tetrahedra.
Moreover, while this article was written with volume data obtained from X-ray tomography in mind,
most of the described algorithms are rather generic and should work well with 
volume data and particle coordinates obtained from magnetic resonance imaging or laser sheet scanning in index matched samples.\cite{stannarius:17,dijksman:12,dijksman:17}.

\section{Visualizing raw data}
\label{sec:visual}

\begin{figure*}[tb]
\begin{center}
    \includegraphics[width=\textwidth]{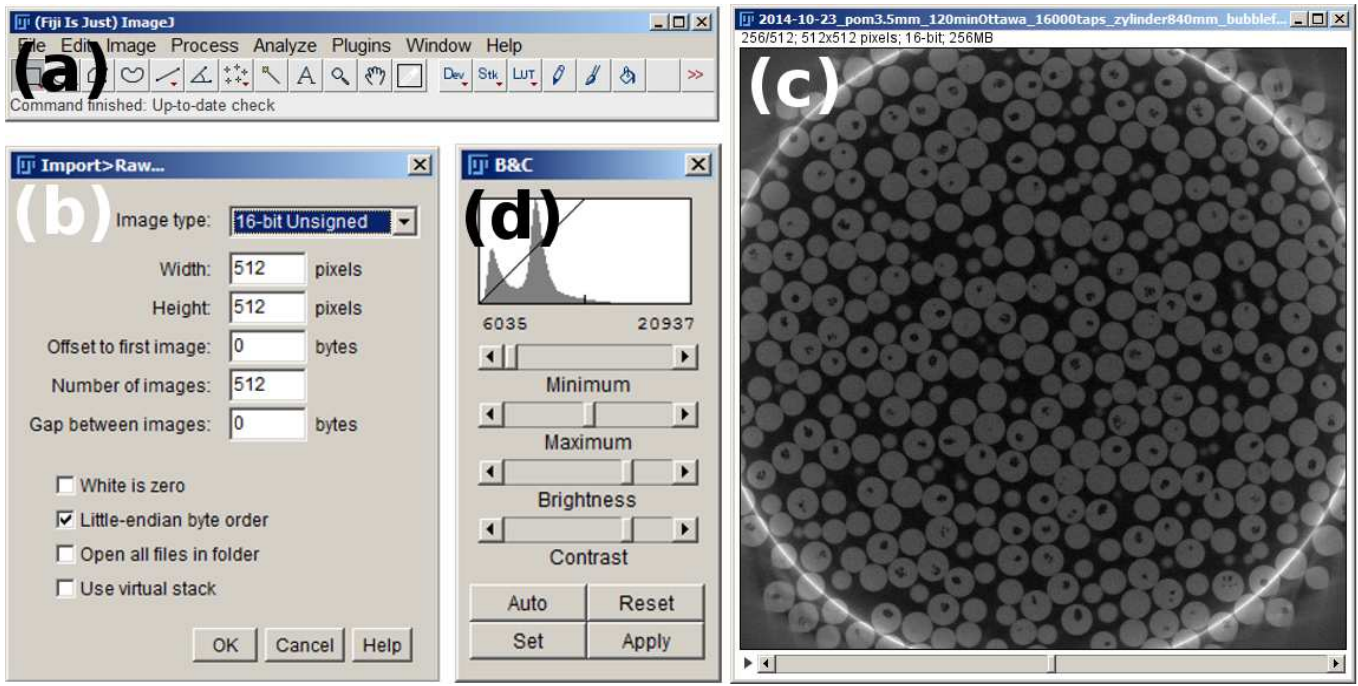}
\end{center}
\caption{
Visualizing raw data in \texttt{Fiji}.
\textbf{(a)} \texttt{Fiji}'s main window after starting the program. 
\textbf{(b)} import raw dialogue, here listing the parameters needed for visualizing the example data. 
\textbf{(c)} one XY slice of the example volume when imported to \texttt{Fiji}. 
\textbf{(d)} the brightness and contrast dialogue allows to adjust the look up table for displaying the image data.
}
    \label{fig:fiji}
\end{figure*}

Every commercial X-ray tomography setup will come with a software which not only allows you to reconstruct
the volume data of your granular packing but also to visualize this volume in a number of ways. However, as this software
is usually a closed-source program, it is a dead-end road for any advanced analysis of the data.
Therefore the first step is to export the data as a three-dimensional volume file, which is typically 
accompanied by a small text file which contains the size information required to read this volume. If the manufacturer
of a 3D imaging setup is not willing to provide such an export option this should be considered a deal-breaker.   

After exporting the volume data set, you might want to inspect it visually and possibly also 
preprocess it. A powerful and open source software for these purposes is \texttt{Fiji}. 
\texttt{Fiji} is written in Java and uses multiple windows to perform its tasks; 
the main window which appears after startup is shown in figure \ref{fig:fiji}(a). 
For more information on how to obtain and run \texttt{Fiji}, see appendix \ref{sec:appfiji}.

To load the volume data, open the import RAW dialogue, shown in figure \ref{fig:fiji}(b) by
selecting \textit{File} $\rightarrow$ \textit{Import} $\rightarrow$ \textit{RAW}  from the menu bar.
Here you will need to provide a number of parameters that can be divided in three groups:  volume specific, visualization choices, and machine specific. 
The most important ones are the volume specific parameters,\textit{Width}, \textit{Height} and \textit{Number of Images} which represent the dimensions of the volume in $x$, $y$, $z$ respectively. 
To open the example dataset accompanying this paper, you will need to insert the values shown in  figure \ref{fig:fiji}(b).
A detailed description of all parameters in the dialogue is given in appendix \ref{sec:appfiji}.

After the import of the raw file is complete, a new window will open, which will show the central XY-slice of the volume. For the example data 
it should look like figure \ref{fig:fiji}(c). The slider at the bottom of the window allows to scan through all slices of the volume.
When moving the mouse over a voxel of the image, the respective coordinates within the image and the voxel's grey value
will be shown in the status bar in {\sc Fiji}'s main window.

To evaluate individual features of an image, it is often necessary to adjust the brightness and contrast of the image.
This can be done by selecting {\it Image} $\rightarrow$ {\it Adjust} $\rightarrow$ {\it Brightness/Contrast} from the menu bar.
The dialogue shown in figure \ref{fig:fiji}(d) will open up, allowing you to tweak the respective values.
The figure in the top part of the window shows the gray value histogram and the look up table (indicated by the straight line)
used in the image window. The latter indicates how the gray value in the image will be mapped onto the gray values on the screen
(represented by the vertical axis). This look up table can be adjusted either
by changing the minimum and maximum sliders or the brightness and contrast sliders.

Often it is also beneficial to look at a different projection of the volume data. 
The default view after loading shows a XY slice through the volume.
To change the projection plane, select {\it Image} $\rightarrow$ {\it Stacks} $\rightarrow$ {\it Reslice}.
Leave all parameters on the default settings and click ok. 
This will create a XZ projection of the volume that will open in a new window.

\section{Finding particles}
\label{sec:find_particles}

The first and most important step in analyzing the volume file is to detect the positions (and for 
non-spherical particles also orientations and size) of all individual particles in the sample. The resulting
list of particle coordinates is the input data for all subsequent analysis steps.  
The difficulty of this image processing step comes from the inherent measurement noise
and imperfections which create ambiguities how to  to assign individual voxels 
to specific particles.\cite{saadatfar:12} 

For monoschematic particles, such as spheres, ellipsoids, cylinders or 
tetrahedra, this problem becomes much simpler, because the a-priori knowledge of their shape 
can be used in particle detection and boundary detection between neighboring particles.

Only very recently, level-set method have been developed \cite{vlahinic:14,kawamoto:16,vlahinic:16}
which allow to reliably identify the positions and orientations of non-monoschematic particles 
such as sand grains.

The choice of the image processing chain will in general depend on the shape of the
particles, the quality and resolution of the raw data and the effort one is willing to take
to optimize accuracy and detection rate. In this paper, we will describe
a rather generic algorithm based on erosion and the Euclidian Distance Map.\cite{schaller:13} 
This algorithm has been successfully used for identifying both spheres and 
ellipsoids.\cite{schaller:15,schaller:15b} All steps of the image processing chain
have been included into a single program, \texttt{volume2positionList}. 
Appendix \ref{sec:apppartdet} describes how to run it with the demo volume data set.
Additionally, the C++ source code can be downloaded from \cite{url_git_v2p} and serve as
a  starting point for code adapted to your specific problem (\texttt{volume2positionList} 
is published under the \texttt{GPL v3} open source license).
However, depending on your volume data, you might need to adapt the parameters of the program or even want to consider other image 
processing approaches, such as a direct watershed transformation \cite{saadatfar:12,cao:13,gillman:13,hurley:16} or 
cross-correlation with template images. \cite{zhao_phd:13,baranau:16} 
 
This section is split into three parts. In section \ref{sec:imageprocessing} we describe the image pre-processing steps 
necessary to obtain a reliably binarized image where all voxels belonging to any of the particles
are set to a grey value of one while the rest is set to zero. Some of these steps  are necessary to compensate 
for deficiencies of x-ray tomography (noise, beam hardening), others are required due to
imperfections of the particles. Then in section \ref{sec:centroidMethod} we describe how to 
assign all white voxels to the correct particle so that their center of mass positions can be computed.
For spherical particles this is equivalent to the particle's center position.
Finally, section \ref{sec:findingtetrahedra} discusses an image processing approach 
suitable for particles with flat faces such as cubes or tetrahedra.

\subsection{Image Pre-processing}
\label{sec:imageprocessing}
\begin{figure*}[t]
    \hspace{-2cm}
    \begin{center}
        \includegraphics[width=\textwidth]{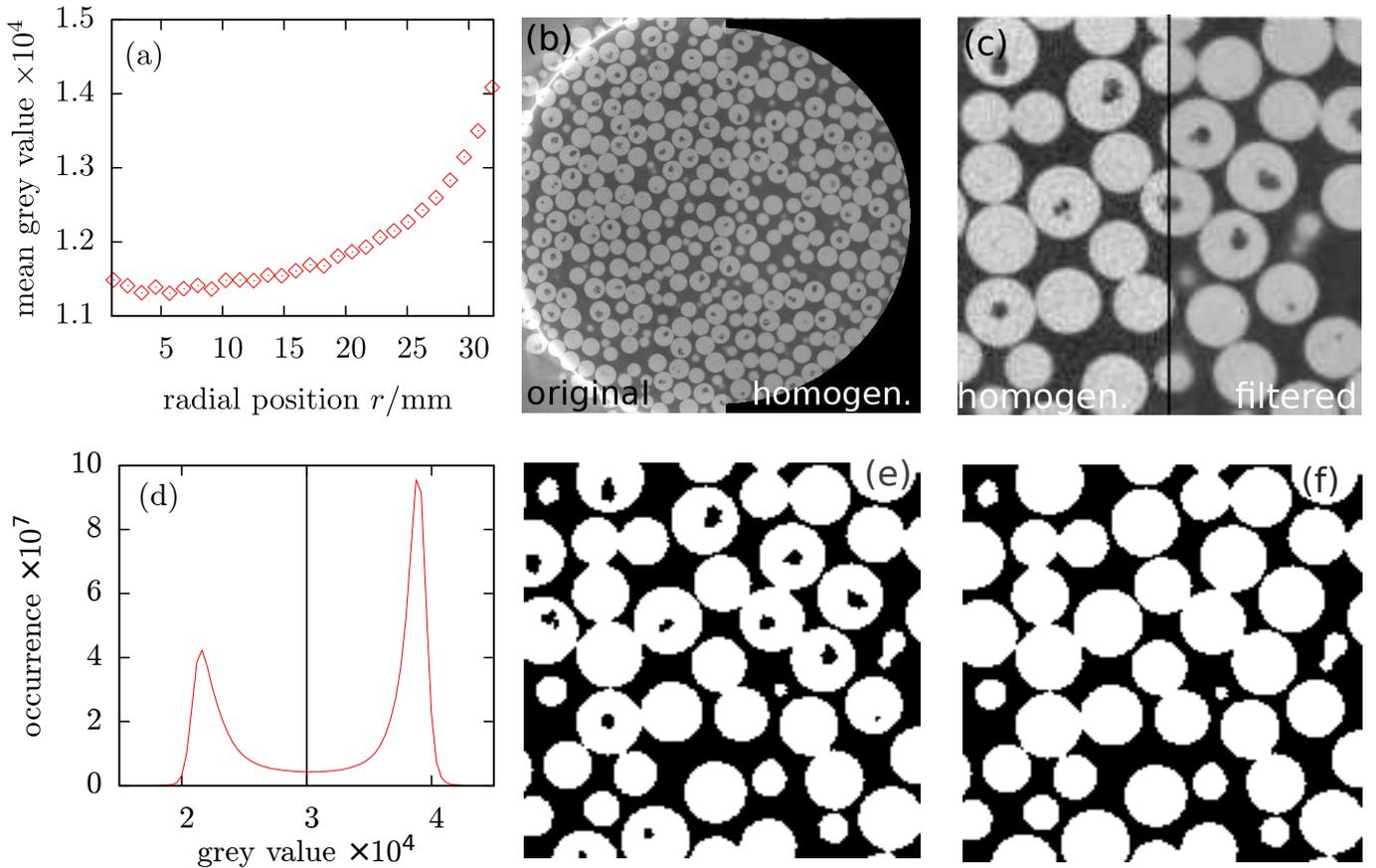}
    \end{center}
    \caption{Pre-processing steps of the example tomogram.
       (a) The azimuthally averaged grey values. Beam hardening leads to increasing grey values with distance from the center axis of the tomogram.
       (b) A tomographic slice. left half: original image data. Right half: after the radial intensity correction. The outer regions are set to black.
       (c) A small area of the homogenized (left half) and filtered (right half) volume. 
       (d) The histogram of all grey values. The left peak correspond to air voxels, the right peak to particle  voxels. The black line is the binarization threshold calculated by Otsu's method.
       (e) A slice of the binarized volume, which shows voids inside the particles.
       (f) Same slice as in panel e but after filling the internal voids.}
    \label{fig:preprocessing}
\end{figure*}

\subsubsection{Radial intensity correction}
Most X-Ray tomograms  are taken with X-rays sources creating a broad spectrum of wave lengths;
the only relevant exception are monochromatic beams at synchrotrons. 
As the absorption coefficient $\mu$ of all materials depends in a nonlinear way on the the wavelength 
of the X-rays, the assumption of an exponential decay of the X-ray intensity with thickness of the material 
is oversimplified.\cite{kleinschmidt:99} A strictly exponential decay is however the basic assumption underlying 
most tomographic reconstruction algorithm; the resulting artifacts in the volume data are called beam hardening.
 
In our example beam hardening leads to a radial gradient in the  grey value distribution 
with brighter voxels at larger radii and lower grey values at the center of the sample.
The effect becomes especially apparent by averaging the grey values in azimuthal rings (in cylindrical coordinates($r$, $\theta$, $z$): average over $z$ and $\theta$) which is plotted in figure \ref{fig:preprocessing}(a). 
This inhomogeneity has to be corrected before the binarization step which is based on a single threshold 
for the whole dataset. 

Figure \ref{fig:preprocessing}(b) shows a slice of the original tomogram on the left side and 
the radially corrected version on the right side.
To obtain the homogeneous grey value distribution, the grey value of each voxel is divided by a correction factor
which is determined from its radial position and an interpolation of the azimuthally averaged grey values shown  
in figure \ref{fig:preprocessing}(a).

The radiograms taken during the acquisition of the the tomogram do not contain sufficient information to
reconstruct the outmost parts of the tomogram correctly. In the left half of figure \ref{fig:preprocessing}(b) this area
corresponds to the bright circle and all parts even further away from the center. This area is
set to black and thus removed from the further image processing and particle detection.

\subsubsection{Bilateral Filter}
An unavoidable consequence of the mathematical principles underlying the reconstruction of  
tomographic images from radiograms is that the signal to noise ratio 
decreases with increasing spatial frequency.\cite{buzug:08}
This inherent noise is especially problematic if a voxel's grey value
 is close to the threshold of the binarization, which is the next step
in the image processing pipeline. 
The noise can then lead to this voxel not being correctly identified.   

A common remedy to reduce the noise is to low-pass filter the raw data with a Gaussian filter.
This filter loops over all voxel positions $\vec{x}$ in the volume file and computes:
\begin{equation}
  f_\textnormal{gaussian}(\vec{x}) = k \;\sum_{\vec{\eta}} \;o(\vec{\eta})\;\cdot g( \vec{x} - \vec{\eta}) 
    \label{eq:gauss}
\end{equation}
where $f_\textnormal{gaussian}(\vec{x})$ and $o(\vec{x})$ are the grey value of the filtered and the original image at position $\vec{x}$  
and $k$ is a normalization factor. The sum runs over all voxel positions $\vec{\eta}$ in a predefined neighborhood of
$\vec{x}$  and $g$ is a Gaussian function representing the geometric distance between the voxels at $\vec{x}$ and $\vec{\eta}$.
The mean of $g$ is 0 and its standard deviation $\sigma_g$  can be used to control  the extent to which
the grey values at $f_\textnormal{Gaussian}(\vec{x})$ are blurred.  

A significant downside of the Gaussian filter is that it does not take the absolute differences between grey 
values into account and is thus blurring the edges in the image. To avoid this effect we apply a
bilateral filter \cite{tomasi1998bilateral} which reduces experimental noise without blurring edges. 
It does so by multiplying the sum with a second Gaussian term $ p(o(\vec{x}) - o(\vec{\eta}))$ (mean 0, standard deviation $\sigma_p$) which represents the 
photometric distance i.e.~the grey value difference between the two voxels under consideration: 
\begin{equation}
    f_\textnormal{bilateral}(\vec{x}) = k(\vec{x})\;\sum_{\vec{\eta}}  \; o(\vec{\eta})\;\cdot g( \vec{x} - \vec{\eta})\;\cdot\; p(o(\vec{x}) - o(\vec{\eta}))
    \label{eq:bilat}
\end{equation}
Figure \ref{fig:preprocessing}(c) shows an enlarged area of the tomogram before (left half) and after (right half) bilateral filtering. 
In general, the standard deviations of the two Gaussian functions, $\sigma_g$ and $\sigma_p$, have to be adapted to the image material at hand. 
For the sample data, choose $\sigma_g = 1.75$ (in units of voxel sidelength) and $\sigma_p = 2000$ (in units of grey values).

\subsubsection{Binarization}
\label{sec:binarize}
The next step in the image processing chain is to create a
binary volume where all voxels belonging to particles are assigned a value of one (white) and
all voxels representing air are assigned a value of zero (black). This
assignment is solely determined by  a grey value threshold.
Figure \ref{fig:preprocessing}(d) shows the grey value histogram of our pre-processed demo volume, 
the lower peak is formed by the air voxels, the higher peak represents voxels from particles.
The optimal threshold, indicated by a vertical line in figure \ref{fig:preprocessing}(d), 
is calculated using \textit{Otsu}'s method:\cite{otsu1975threshold} it minimizes the weighted sum 
of the standard deviations of the two populations created by this threshold.  
Figure \ref{fig:preprocessing}(e) displays the same area as figure \ref{fig:preprocessing}(c)
but after binarization.

\subsubsection{Filling internal voids}
A frequent imperfection in commercially available granular particles are 
cavities created during the manufacturing process, especially but not exclusively 
during injection molding. In order to not obstruct the 
calculation of the centroid of each particle, these voids have to be removed. 
If the voids are completely internal, i.e.~they are not connected to the exterior
of the particle, they can easily be corrected by the following algorithm:
Using the Hoshen Kopelman algorithm described in appendix \ref{sec:hk}, all clusters of black
voxels are identified. The largest of these clusters is the air surrounding all particles,
all other clusters are cavities inside of particles. 
By setting the voxels of all but the largest clusters to 1, these cavities will be filled. 
The result of this operation is shown in figure \ref{fig:preprocessing}(f).

In case the voids are connected to the outside of the particle, an improvement of the 
raw data can still be obtained by applying above algorithm to the individual slices in alternating
orientations.\cite{zhao_phd:13}


\subsection{Identifying particles using the Euclidean Distance Map}
\label{sec:centroidMethod}

\begin{figure}[tb]
\begin{center}
    \includegraphics[width=0.5\textwidth]{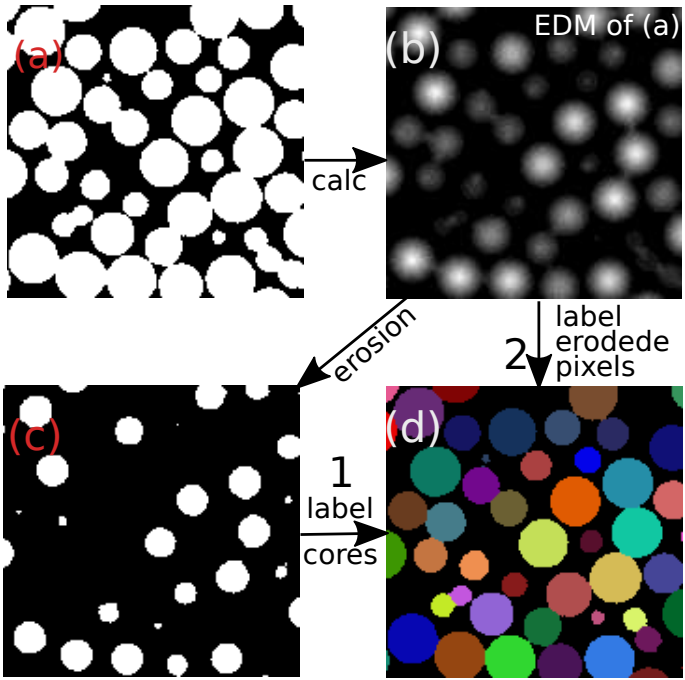}
\end{center}
\caption{Identifying particles by assigning the white voxels using the Euclidean Distance Map.
   (a) Slice of the binarized volume after the voids inside the particles have been filled.
   (b) Euclidean Distance Map (EDM). The voxel's grey value corresponds to it's distance to the nearest black voxel. 
   (c) the eroded volume,  separating the particle cores.
   (d) the processed volume, different colors correspond to different particle IDs.
}
    \label{fig:centroids}
\end{figure}

Purpose of this step is to assign to each white voxel, i.e.~voxel belonging to a particle,   
a number (the particle ID) that identifies to which particle this voxel belongs. Then the 
position of each particle can be computed as the centroid of all its voxels. 

This is done by first computing the \textit{Euclidean Distance Map} (EDM). 
In the EDM each white voxel is assigned the geometric distance (in units of voxels) to the closest black voxel.  
Thus all black (air) voxels are assigned 0. In consequence, the center voxels of the particles 
are assigned with the highest values, as they are farthest away from any black voxel.
Figure \ref{fig:centroids}(b) shows the EDM of the binarized image in figure \ref{fig:centroids}(a).

The EDM is then the starting point of a two step algorithm. 
In a first step, the particles are separated using image erosion. The remaining separated particle cores are labeled. 
In a second step all voxels, which have been eroded in step one, are assigned to their respective particle cores.

After binarization all particles form, due to their contacts, one connected white blob. 
The erosion step separates this blob into individual particles by removing an outer layers of thickness $\lambda$. 
Practically, this is done by thresholding the EDM with the chosen erosion depth $\lambda$; 
all voxels with an EDM value smaller than $\lambda$ will be set to black, the remaining to white.
Figure \ref{fig:centroids}(c) show the erosion result of our demo slice with $\lambda$ = 5.  
Similar to the two $\sigma$ values of the bilateral filter, the value of $\lambda$ has to be 
adapted to the respective volume data. If $\lambda$ is too small, it will not remove all the connections 
between individual particle clusters. If $\lambda$ is too large, particles will be eroded completely away.
Finally, the voxels in the now separated particle cores are assigned 
an unique label, the particle ID, using  the Hoshen Kopelman algorithm described in 
appendix \ref{sec:hk} (this time working with the white voxels).

In the second step all voxels, that have been eroded in the first step, have to be assigned their correct particle ID.  
This is achieved by performing an an uphill gradient search on the EDM which will point for each voxel to the closest 
particle center.  Figure \ref{fig:centroids}(d) shows the labeled version of our demos slice with 
each color representing another label.

The labeled volume contains also clusters of particles that touch the radial cut that has been done during 
image pre-processing, as well as particles that touch the bottom and top of the volume.
Those clusters have to be removed from the labeled volume, because they do not correspond to 
completely  detected particles. This is done by removing all clusters from the labeled image, which are 
closer than one particle diameter to the radial boundary or that touch the bottom or the top slice of the tomogram.
If the analysis boundary coincides with a smooth container boundary, strong layering effects will occur 
\cite{zhang:06,jerkins:08,desmond:09} and the subsequent analysis should ignore rather 
three to four of the outmost layers. 

The labeled image can now be used to calculate the centroid of each particle by averaging the coordinates 
of all voxels with the same label. The precision of the centroid will depend on the accuracy
of the detection of the boundary voxels of the particle. However, due to the large number of boundary voxels, 
typically sub-voxel accuracy can be expected. A method to measure this accuracy will be discussed in 
section \ref{sec:pair_correlation}. Additionally, counting the number of voxels of each label will yield the volume 
of the respective particle. The list of particles centers, volumes and IDs is the main result of the 
image processing chain.

After removing the only partially detected particles at the boundaries, the volume distribution of the remaining
particles should be rather narrow peaked, reflecting the narrow size distribution of the particles as well as the imaging system resolution. The presence of particles with roughly two or three times 
the average particle volume indicates that the erosion depth $\lambda$ has been chosen to small. 

The algorithm discussed in this section works well for monodisperse spheres and 
ellipsoids.\cite{schaller:15,schaller:15b} For ellipsoids not only the center of mass, but also
lengths and orientation of each ellipsoid's axis have to be determined. This can be done by
computing the equivalent of the moment of inertia tensor (as described in 
section \ref{sec:karambola}) of all voxels belonging to an ellipsoid.
In general, the algorithm described above can be expected to be suitable for all types of 
regular particles which are neither concave nor posses flat faces. 
A method handling particles, which can form  face to face contacts, will be discussed in the next subsection.


\subsection{Finding tetrahedra}
\label{sec:findingtetrahedra}

\begin{figure}[ht]
\begin{center}
    \includegraphics[width=0.5\textwidth]{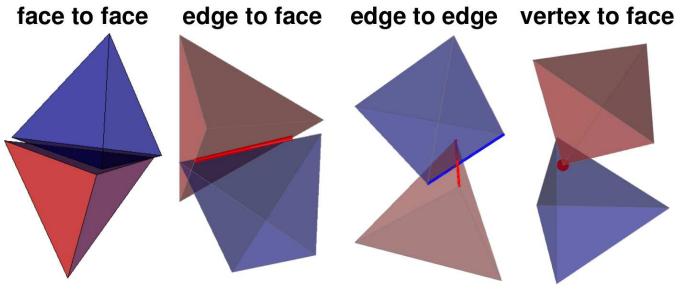}
\end{center}
\caption{The four different contact types found in tetrahedra packings.\cite{neudecker:13} Reproduced with permission from Phys. Rev. Lett. 111, 028001 (2013). 
Copyright 2013 APS. 
 }
    \label{fig:tetra_contact}
\end{figure}

Many granular materials such as salt or sugar are composed of constituents which are created by
a crystallization process. The resulting flat faces of the particles quadruples the number of ways 
particles can form contacts (compared to the point-like contacts between spheres). 
This is demonstrated in Figure \ref{fig:tetra_contact} for the example 
of two tetrahedra: there are face to face, edge to face, edge to edge, and vertex to face contacts.

The mechanical differences resulting from the different types of contacts will be discussed in 
section \ref{sec:contacts}. But the presence of face to face contacts is also a challenge 
for the development of particle detection algorithms because two tetrahedra forming a well 
aligned face to face have a joint minimum in the Euclidian Distance Map. Which implies
that they can neither be separated by the erosion step described above nor a watershed 
transformation. Therefore an alternative approach to the particle detection problem is needed.

\begin{figure*}[t]
\begin{center}
    \includegraphics[width=\textwidth]{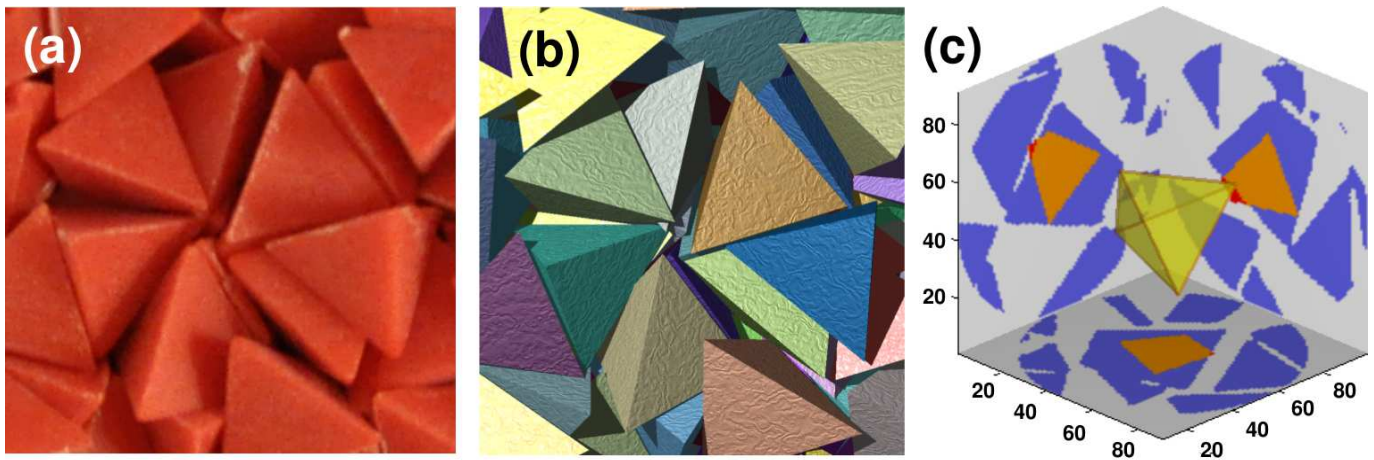}
\end{center}
\caption{Detecting tetrahedra. 
 (a) Photography of packing surface.  The tetrahedra are made of polypropylene  and have a side length of 7mm. 
 (b) Rendering of the particles detected in the same sample.
 (c) Visualizing the steepest ascend algorithm. The three sidewalls of the cube are orthogonal cuts through the 
 volume with gray corresponding to the empty space between the (blue) particles. The yellow tetrahedron in the center 
 represents the probe body which is translated, rotated, and grown in order to maximize the overlap (indicated by orange)
 with voxels belonging to particles and minimize the overlap with air voxels (shown in red).\cite{neudecker:13} 
Reproduced with permission from Phys. Rev. Lett. 111, 028001 (2013). Copyright 2013 APS. 
}
    \label{fig:tetra_detect}
\end{figure*}

Figure \ref{fig:tetra_detect} (a) and (b) depict a photograph of real tetrahedra as well as a rendering of the same tetrahedra after X-ray tomography and particle detection, thus  demonstrating that the detection algorithm 
described in reference \cite{neudecker:13} works well for tetrahedra, resulting in 
detection rates larger than 99.8 \%. The algorithm is based on a steepest ascend 
which is explained in figure \ref{fig:tetra_detect} c: After preprocessing steps similar to the ones described
in section \ref{sec:imageprocessing}, a tetrahedral probe body is 
translated, rotated and grown within the binarized sample  in such a way that its overlap with 
voxels belonging to tetrahedral material is maximized.  
When the size of the probe body has reached the dimensions of an actual particle, a new
tetrahedron has been found. After determining its center of mass and orientation, all voxels belonging 
to this tetrahedron will be set to black in order to facilitate the detection of the remaining particles.
More details on the algorithm can be found in.\cite{neudecker_phd:13}

\section{Pair correlation function}
\label{sec:pair_correlation}

\begin{figure*}[t]
\begin{center}
    \includegraphics[width=\textwidth]{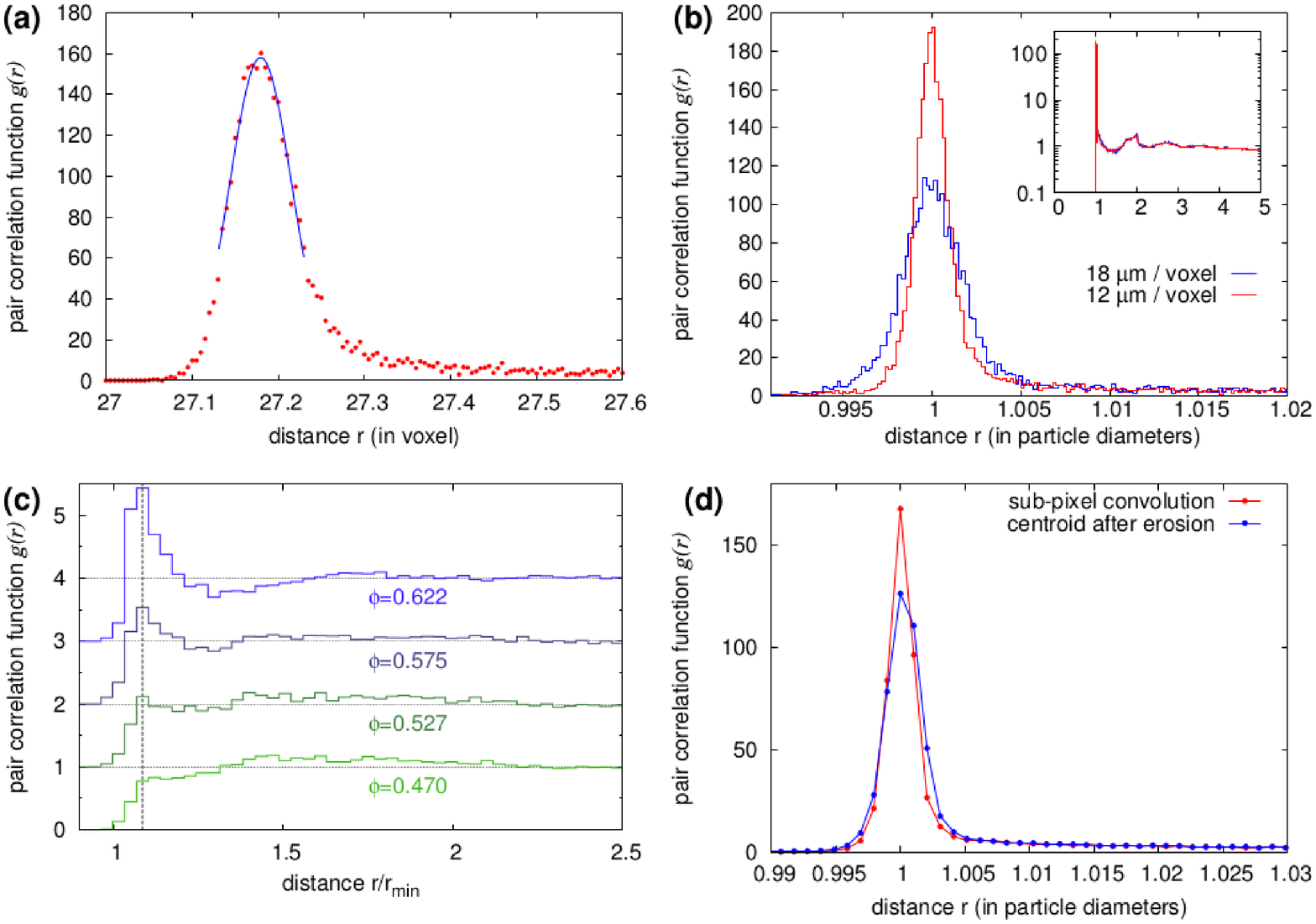}
\end{center}
    \caption{Information provided by the first peak of the pair correlation function.  
    (a) A fit of the first peak of the pair correlation function with a Gaussian (blue line) provides a precise
    estimate of the average diameter of the spheres. Data points are from the analysis of the demo data set accompanying this paper.
    (b) The influence of the spatial resolution of the tomograms can be demonstrated with packings of highly monodisperse spheres
    ($d$ = 351 $\pm$ 0.5 $\mu$m). Figure reproduced with permission from Soft Matter 12, 3991 (2016). Copyright 2016 Royal Socienty of Chemistry.
    (c) For tetrahedra the shortest possible distance $r_\textnormal{min}$, which is approximately 0.408 times the side-length, is
    not the most likely distance between individual particles (indicated by a vertical dashed line). Offsets have been 
    added for improved visibility.
    Reproduced with permission from Phys. Rev. Lett. 111, 028001 (2013). Copyright 2013 APS.
    (d) Datasets of sufficiently monodisperse beads ($d$ = 351 $\pm$ 0.5 $\mu$m) do also allow a comparison of different
    particle detection algorithm.\cite{zhao_phd:13}
 }
    \label{fig:pair_correlation}
\end{figure*}

Typically, the first analysis step done after finding the particle coordinates is to compute the pair correlation
function $g(r)$ (often also called radial distribution function). 
$g(r)$ describes how the average number density changes  while moving away a distance $r$ from an (arbitrary) reference particle. 
The two reasons for $g(r)$ being one of the first steps in most analysis is that it not only gives some insight in the structure 
of the packing but also provides a measure for the quality of the image acquisition and particle detection.

A discretized version of the pair correlation function can be computed by counting the number of particles in spherical
shells around a reference particle: 
\begin{equation}
    g(r, \Delta r) =  \frac{\langle N(r + \Delta r) - N (r)\rangle}  {V_\textnormal{shell} \rho}
    \label{ref:eq_gofr}
\end{equation}
$N(r)$ is the number of particles within a sphere of radius $r$ around a given particle.
$N(r + \Delta r) - N (r)$ corresponds to the number of particles in a spherical shell
of radius $r$ and thickness $\Delta r$ (we assume here $\Delta r$ to be small compared to $r$). 
The average $\langle ... \rangle$ runs over all
particles in the sample. In practise, the numerator is computed by calculating the 
center to center distance of all pairs of particles and then storing this information in a histogram.

The  normalisation in the denominator consists of two parts. First the 
volume of the spherical shell  $V_\textnormal{shell}$ does depend on $r$: 
$V_\textnormal{shell} = 4 \pi r^2 \Delta r$.  Secondly, by dividing with 
the number density $\rho = \frac{N_\textnormal{tot}}{V_\textnormal{tot}}$ 
we assure that an uncorrelated system will have $g(r) = 1$.

In analyzing experimental data, $\Delta r$ has to be chosen small enough to not smear out important features,
but large enough to have good statistics in each bin. For the sample data $\Delta r = 0.005$\,voxel works well.
Another effect to be taken into account is the finite size of all experimental samples.
The larger $r$ becomes, the larger will be the number of particles for which at least a part of the spherical shell 
will be outside the sample volume and therefore empty. Consequentially, the experimentally determined $g(r)$ decays to zero  
for large distances $r$.

In an ideal gas, or any other system without correlations, $g(r)$ is a constant and set to one by the normalization with the average density.   
In an amorphous packing of monodisperse spheres, $g(r)$ will appear as in the inset of figure \ref{fig:pair_correlation}b.
For distances shorter than a particle diameter $d$ the value of $g(r)$ is zero, because the spheres cannot overlap. 
Finite values in this range point to either the presence of additional smaller particles or problems 
during image processing. At the distance of one particle diameter there is a strong peak, which is formed by all 
the pairs of particles which are in contact. Due to both the polydispersity of the particles and 
noise in the experimental data, this peaks will be broadened. This effect which can e.g.~be seen in the main plot
in figure \ref{fig:pair_correlation} (b) will be discussed in more detail below. In amorphous systems without far-reaching order
there are other small peaks of $g(r)$ for the values of $\sqrt{2}d$, $2d$, and $\sqrt{3}d$, which is shown in the inset of \ref{fig:pair_correlation} (b). 
But for larger distances $g(r)$ approaches the value of one characteristic for disordered systems (aside from the finite size 
effects discussed above).

$g(r)$ does provide important insight into structural differences between packings of spheres \cite{seidler:00,aste:04,aste:05,aste:05a}
rods,\cite{yadav:13} ellipsoids,\cite{xia:14a} tetrahedra,\cite{neudecker:13} or granular chains.\cite{zou:09} 
Figure \ref{fig:pair_correlation}c demonstrates for example that in packings of tetrahedra the closest possible distance
is {\it not} the most likely contact configuration.\cite{neudecker:13}
This effect is a consequence of the lower probability of a perfect 
face-to-face alignment compared to either slightly shifted face-to-face or low angle face-to-edge contacts. 
Within the jamming paradigm,\cite{vhecke:10,liu:10} the shape of the first peak of $g(r)$ is used to derive 
how the number of contacts in a packing of compressible spheres will change with pressure. 
Figure \ref{fig:pair_correlation}c asserts that these scaling laws 
will not be applicable to packings of soft tetrahedra. 

In an ideal world of absolutely monodisperse, hard spheres and zero error in the coordinate detection, the left shoulder
of  the first peak of the pair correlation function would be a step function. The shape of the right shoulder 
will reflect the extent to which particles have almost formed contacts. For frictional particles the exact analytic
form of this decay is not known, but the results discussed below indicate that it will also be a steep decay.
In praxis any broadening of the first peak can therefore be traced back to the polydispersity of the spheres and/or
experimental noise.\footnote{In a system of soft particles, this broadening might also be due to deformations of the particles.\cite{dijksman:17}} The latter is usually well modeled by a Gaussian distribution, the former can often be approximated by a 
Gaussian. 

Figure \ref{fig:pair_correlation}a shows a Gaussian fit to the first peak in the $g(r)$ of our demo dataset.
The two fit parameters are the mean of 27.177 voxels  and the standard deviation $\sigma$ of 0.036 voxels.
The mean value corresponds to the interpolated maximum of $g(r)$, which is the most frequent separation two sphere centers will have. 
It is therefore our best estimate for the  mean diameter of the spheres $d_\textnormal{mean}$. 
The quality of this method to determine average particle diameters is demonstrated in   
a study of segregation in a polydisperse system \cite{finger:15} where $d_\textnormal{mean}$ changes
of fractions of a percent can be detected reliably. 

The standard deviation $\sigma$ of the Gaussian fit is a convolution of the effects of polydispersity, 
experimental noise and particle detection. If the experimental data are taken with sufficiently monodisperse spheres, $\sigma$
can be used to characterize the quality of the image acquisition (figure \ref{fig:pair_correlation} (b),  
taken from,\cite{baranau:16}) compare the quality of different sphere detection algorithms 
(figure \ref{fig:pair_correlation} (d),  taken from \cite{zhao_phd:13}), or even compare 
different imaging methods (X-ray versus neutron tomography \cite{murison:15}). 

Pair correlation functions for the example data can be calculated with \texttt{raps},\cite{url_git_raps} 
which can compute also a number of other structural characteristics of sphere 
packings.
\texttt{raps} is licensed under the LGPL and is free to use, further details 
can be found in  appendix \ref{sec:appraps}. The fit to the first maximum of $g(r)$
shown in figure \ref{fig:pair_correlation} (a)  can be performed a with 
script described in appendix \ref{sec:appcns}.

\section{Determining contact numbers}
\label{sec:contacts}
For a granular packing to be mechanically stable, its particles needs to have enough contacts $Z$ to block all their translational  and rotational 
degrees of freedom. The minimum number of contacts necessary to achieve this is the so called isostatic contact number 
$Z_\textnormal{iso}$. For frictional spheres\cite{vhecke:10} $Z_\textnormal{iso}$ equals four. 
Determining $Z$ has always been a desired experimental goal. Before the advent of tomographic imaging, researchers had mixed
particles with small amounts of paint which was then attracted to the contact points by capillary forces. After the paint had dried, 
the packing could be disassembled and the analysis of the paint marks at the surface allowed to estimate $Z$ for packings of 
spheres \cite{bernal:60,pinson:98} and cylinders.\cite{blouwolff:06}

The availability of tomographic images seems to allow the direct observation of contacts in form of connected 
pathways of voxels between particle centers. However, in practice this idea is hard to realize.\cite{saadatfar:12, hurley:16}
The actual contact between hard particles is formed in a small area only, 
corresponding to a small number of voxels. Even a minor error in the choice of   
the binarization threshold can erase or fill erroneous voxels and therefore  
make the detection going completely wrong.

An alternative approach is to a) not consider contacts between individual particles but to compute 
$Z$ from the whole ensemble of particles at once using their center of mass coordinates. 
And b) to rely on the information contained in the whole surface of the particles, not only at the position of potential 
contacts. The second point is already used in our particle finding algorithm: small mistakes in the binarization threshold 
will influence all voxels at the particle boundary the same way. The resulting  over-erosion or dilation is radially symmetric 
and will therefore have only a small influence on the determined center of mass. 

Point a) is exploited by using the information from the ensemble of all particles to first determine the best estimate for the particle diameter itself 
(as discussed in section \ref{sec:pair_correlation}) and then to fit a model to the data which allows to
identify both $Z$ and the strength of experimental noise and polydispersity. This ensemble based fitting method is described in more detail in 
the next subsection. 
The basic idea for this approach has first been suggested and applied by.\cite{aste:05,aste:05a} In the last years
the method has then been used to determine $Z$ for packings of  
spheres and ellipsoids,\cite{schaller:13,schaller:15} tetrahedra,\cite{neudecker:13} and cylinders.\cite{zhang:14}

\subsection{Ensemble based fitting}
\label{sec:cns}
\begin{figure*}[t]
\begin{center}
    \includegraphics[width=\textwidth]{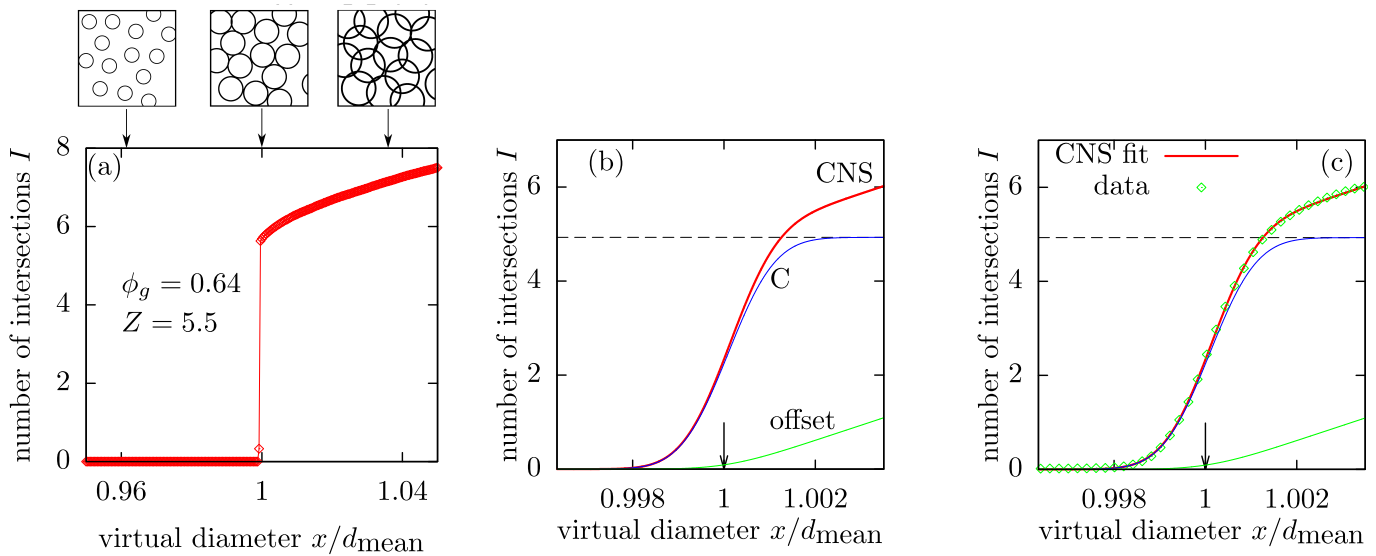}
\end{center}
\caption{Determining Contact numbers by fitting the Contact Number Scaling function to the experimental data.
(a) The number of intersections between spheres $I(x)$ in an ideal (monodisperse, noise free) world. 
The three images at the top show  two-dimensional schematics for different virtual side lengths $x$.
(b) The CNS function and its components.
(c) A fit of the CNS function to experimental data.
}
    \label{fig:cns_function}
\end{figure*}

This approach to determine the global contact number $Z$ works on the basis of the information contained in the first peak of $g(r)$. 
The first step is to endow all particle centers in the packing with a virtual diameter $x$ and then determine the average contact number $I(x)$ 
of this packing by counting the number of intersections between these virtual spheres.
For monodisperse spheres, $I(x)$ is equal to an integral over $g(r)$ up to the value $x$. 

Figure \ref{fig:cns_function}(a) depicts $I(x)$ for an idealized dataset of monodisperse spheres 
and in the absence of experimental noise. As the data in figure 
 \ref{fig:cns_function}(a) has been extracted from a packing created by simulation,\cite{makse-rcpdata} these conditions
are indeed fulfilled. For $x<d_\textnormal{mean}$ there are no intersections at all, thus $I(x)$ is zero.
At the actual sphere diameter $d_\textnormal{mean}$, $I$ jumps to the global contact number $Z$. 
And for larger values of $x$, $I$ keeps increasing due to the formation of spurious contacts. Modeling
this behavior results in a function $Z\, \cdot  \theta(x-d_\textnormal{mean}) \,+\, \textnormal{off}^\textnormal{ ideal}(x)$ 
with $\theta$ being the Heaviside step function and $\textnormal{of}^\textnormal{ ideal}(x)$ modeling the a priori unknown
increase of $Z$ above $d_\textnormal{mean}$. 

In any  experimental system, the particles are not ideally monodisperse and there will be experimental noise 
in  the imaging system. These effects are similar to convoluting the above described ideal model with a Gaussian function. 
The resulting model is called Contact Number Scaling (CNS) function: 
\begin{equation}
    \begin{split}
        \textnormal{CNS}(x) &= C(x) + \textnormal{off}^\textnormal{ real}(x)\\
        C(x) &= Z/2 \cdot \left( \textnormal{erf}\left( \sigma\cdot (x-d_\textnormal{mean}) \right)+1 \right)
    \end{split}
    \label{eq:cnsfunction}
\end{equation}
with $\sigma$ being the width of the error function. 
Examples of CNS($x$), $C$($x$), and $\textnormal{off}^\textnormal{ real}(x)$ are shown in figure \ref{fig:cns_function}(b).

By fitting equation \ref{eq:cnsfunction} to the experimental $I(x)$ as shown in figure \ref{fig:cns_function}(c) , 
the global contact number $Z$ can be determined.
Here $d_\textnormal{mean}$ is not a fit parameter as it has already been determined from the Gaussian fit to $g(r)$
discussed in section \ref{sec:pair_correlation}. If $Z$ is determined for a larger series of tomographies all taken with
identical imaging conditions, the number of effective fit parameters can be reduced further as neither the 
polydispersity nor the experimental noise will depend on the individual experiment. Consequentially, $\sigma$ 
should also be the same for all experiments. We can therefore perform a second round of fits 
where $\sigma$ is held constant at a value which is the mean of the first round of fits.\cite{neudecker:13} 

$I(x)$ can be computed using \texttt{raps}. It is then stored in a file named \texttt{cns\_example.dat}.  
A detailed description of the fit procedure can be found in appendix \ref{sec:appcns}; 
a minimal fit script based on the open source program \texttt{gnuplot} \cite{url_gnuplot} is part of the supplemental material.

Once the global contact number $Z$ has been measured, we can also identify a local contact number $Z_l$ as the overlap between 
individual particles after their size has been rescaled such that the average of $Z_l$ is identical to $Z$. Especially for frictional 
particles, $Z_l$ shows a strong dependence on the local volume fraction\cite{schaller:15,thyagu:15}
which will be introduced in section \ref{sec:local_phi}.  
For particles with different types of contacts, such as tetrahedra, a further step of image 
processing is required to determine the type of contact from the angle between the surface normals of the involved faces.\cite{neudecker:13}
Finally, a number of other measures, such as e.g.~fabric anisotropy,\cite{radjai:09} require knowledge about the position of the individual 
contacts between particles.

\section{Voronoi volumes}
\label{sec:voronoi}

\begin{figure*}[tbp]
\begin{center}
    \includegraphics[width=\textwidth]{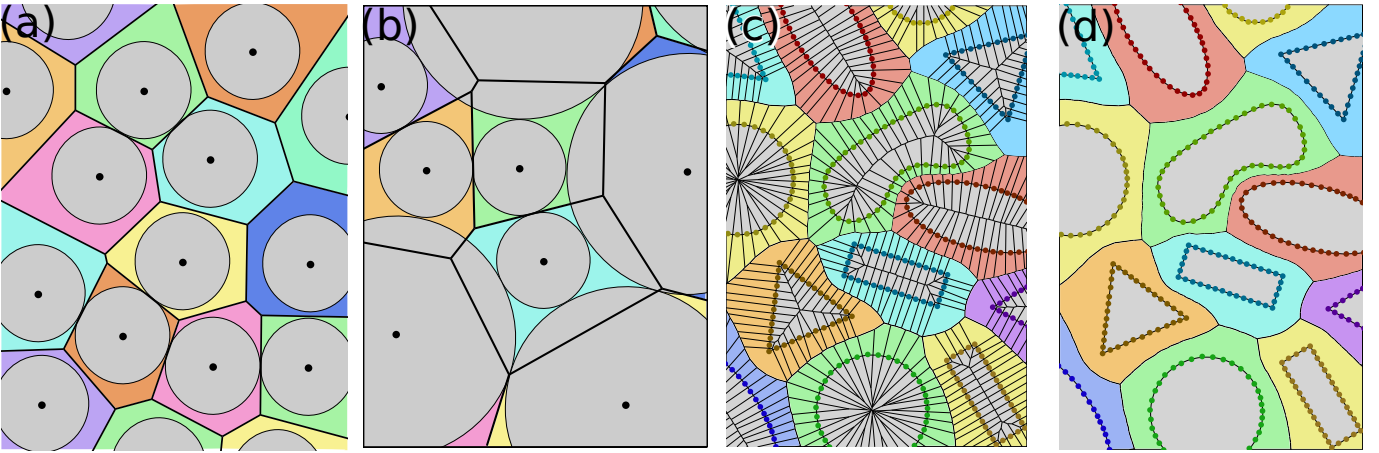}
\end{center}
    \caption{
        Two-dimensional Voronoi and Set Voronoi diagrams.
       (a) A Voronoi tessellation of monodisperse discs.
       (b) Voronoi diagrams are not well suited for a system of polydisperse discs. 
       (c) The first two steps of calculating a Set Voronoi Diagram. 
       First points are spawned on the surface of the particles. Then the classical
       Voronoi diagram for those points is calculated.
      (d) The Set Voronoi cells are calculated by merging all Voronoi cells that belong to the same particle. 
    }
    \label{fig:voronoi_2s}
\end{figure*}

Any attempt to understand granular physics based on the behavior of its individual constituents will need to 
provide a definition of what is considered the local environment of a particle.\cite{blumenfeld:03,song:08,puckett:11,baule:13,baule:14,xia:14} 
For monodisperse spheres the by far most common definition is provided by the Voronoi tessellation: 
the space in the sample is divided into cells such that each cell contains all points that are closer to the center 
of a given particle than to to any other particle center. Figure \ref{fig:voronoi_2s}(a) illustrates this idea
using a two-dimensional packing of monodisperse discs, figure \ref{fig:voronoi_3d}(a) shows the Voronoi cells 
of a sphere packing.  
Unlike some of the competing partitioning schemes,\cite{pica-ciamarra:07} 
it is well defined in both two and three dimensions and it is space-conserving. 
The latter allows us to define a local volume fraction $\phi_l$.

Voronoi diagrams  have a number of geometrically and physically interesting properties. 
For example, the Voronoi volume distribution of sphere packings is well described by Gamma-distributions
\cite{starr2002we, kumaran:05,aste:07,aste:08} and independent from the 
experimental or numerical protocol the packing was created with.
This result has also been generalized to cuboidal granular media.\cite{shepherd:12} 

\subsection{Local volume fraction}
\label{sec:local_phi}
The local volume fraction assigns each individual particle a local density. It is 
defined as 
\begin{equation}
  \label{eq:phi_l}
  \phi_l = \frac{V_p}{V_c}
\end{equation}
with $V_p$ the volume of the particle and $V_c$ the volume of it's cell.

Local volume fractions provide a convenient way to compute the global packing fraction $\phi_g$
from a subset of all particles, thereby e.g.~avoiding systematic errors in the $\phi_g$ values due 
to ordered layers and inaccessible space at container boundaries.\cite{jerkins:08,desmond:09} 
However, while volumes are additive, volume fractions are not. 
Therefore the harmonic mean has to be used to calculate $\phi_g$  from the local packing fractions
of the individual cells $\phi_l^i$:
\begin{equation}
     \phi_g \,=\, \frac{N V_p}{\sum\limits_{i}\limits^{N} V_c^i} 
            \,=\, \frac{N}{\sum\limits_{i}\limits^{N} \frac{V_c^i}{V_p}} 
             \,=\, \frac{N}{\sum\limits_{i}\limits^{N} \frac{1}{\phi_l^i}}
\end{equation}
with $V_c^i$ being the volume of the \textit{i}th Voronoi cell 
and $N$ the total number of particles.

\begin{figure}[tbp]
\begin{center}
    \includegraphics[width=0.5\textwidth]{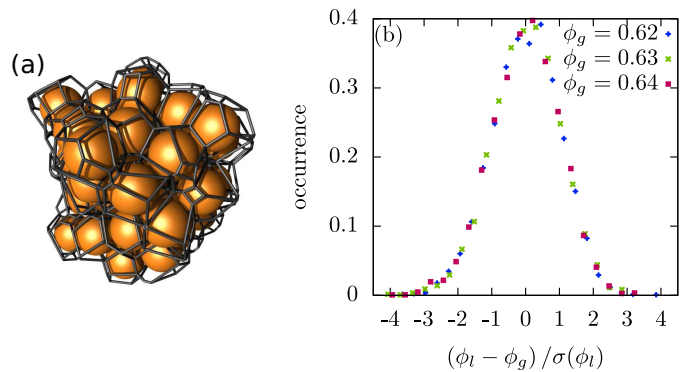}
\end{center}
\caption{Voronoi tessellation in three dimensions.
   (a) A subset of a packing of spheres with its Voronoi cells.
   (b) Rescaled distribution of local packing fractions from systems of monodisperse spheres at different global packing fractions $\phi_g$.
 }
    \label{fig:voronoi_3d}
\end{figure}

Figure \ref{fig:voronoi_3d}(b) demonstrates an interesting property of the local volume fraction
distribution. Rescaling it with:
\begin{equation}
  \label{eq:scaling}
  \phi_l^` = \frac{\phi_l - \phi_g}{\sigma(\phi_l)},
\end{equation}
where $\sigma(\phi_l)$ is the standard deviation of the distribution, 
results in a data collapse for packings taken at different values of $\phi_g$.
Even more unexpected is that this universality extends also to packings  
of oblate ellipsoids.\cite{schaller:15,schaller:15b}
Additionally, it can be shown that contact numbers in packings of spheres and ellipsoids
are best explained by a theory starting from $\phi_l$.\cite{schaller:15}

The local volume fractions of our demo data set can be computed using the program 
\texttt{pomelo}.
Shape analysis of the Voronoi cells can be performed with the program \texttt{karambola}. Both programs will be introduced in the remainder of this section. 

\subsection{Set Voronoi tessellation}
\label{sec:pomelo}
The normal Voronoi tessellation described above is not reliable for packings made from anything else than monodisperse spheres
or disks. The underlying problem is that a single point, the center of mass, is not sufficient to describe the spatial extension 
of the particle. Figure \ref{fig:voronoi_2s}b demonstrates this for a packing of bidisperse disks:
the volume of some disks does not belong to the Voronoi cell of that disk.

This problem is avoided  by a generalization of the Voronoi tessellation, the \textit{Set Voronoi} diagram 
(also called navigation map  or tessellation by zone of influence).\cite{luchnikov1999voronoi,preteux1992watershed}
The Set Voronoi diagram uses the surface of the particles to assign the free space in between the particles.
Thus each Set Voronoi cell contains the volume that is closer to the surface of its central particle 
than to any other particle surface.

Set Voronoi diagrams are usually numerically approximated. First the surface of the particles is 
discretized by spawning a grid of points on it. Then the packing is tessellated by 
calculating the classical Voronoi diagram for all surface points. These two steps are illustrated in 
figure \ref{fig:voronoi_2s}c for a two-dimensional system of arbitrarily shaped particles.
The final Set Voronoi diagram is then obtained by merging all Voronoi cells that belong to the same particle
as shown in figure \ref{fig:voronoi_2s}(d). 

As discussed in section \ref{sec:pair_correlation}, experimental noise will lead to 
inaccuracies in the detected particle coordinates which will result in small overlaps 
between the particles surfaces. As this would lead to erroneous results during the calculation
of the Set Voronoi diagram, all particles have to be shrinked sufficiently in size to resolve those overlaps.
This step will influence for most particles the result of the tessellation; the amount 
of shrinkage should therefore be kept to a minimum.

\subsubsection{Computing Set Voronoi diagrams}

You can use the program \texttt{pomelo} \cite{url_git_pomelo} to calculate both Voronoi and Set Voronoi Diagrams.
\texttt{pomelo} offers built-in functionality for common particle shapes, such as mono- and polydisperse spheres, ellipsoids, 
tetrahedra, and spherocylinders.  Additionally, it offers a generic mode which allows to handle almost any particle shape. 
The command required to run \texttt{pomelo} on the example data is explained in appendix \ref{sec:apppomelo}.

While \texttt{pomelo} does compute the Set Voronoi cells and their volume, it does not characterize their shape geometrically.   
This task is handled by the program \texttt{karambola} discussed in the next subsection. 
Furthermore, in order to limit the size of the Voronoi cells of the outmost layers of particles detected, 
\texttt{pomelo} assumes that the sample is embedded in a rectangular box. This leads to 
erroneous large Voronoi cell volumes in these outer layers of the sample which have then to be discarded.

\subsection{Describing Voronoi cells with Minkowski Tensors}
\label{sec:karambola}
The best way  to quantify the shape and morphological properties of the (Set-) Voronoi cells is by computing
their \textit{Minkowski Tensors}.\cite{schroeder-turk:10,turk2013karambola} 
The importance of the Minkowski tensors is underscored by Alesker's theorem, which states 
any additive, motion-covariant, and continuous function
can be described by a linear combination of these six different tensors.\cite{alesker1999description} 

\begin{figure}[tbp]
    \centering
    \includegraphics[width=0.5\textwidth]{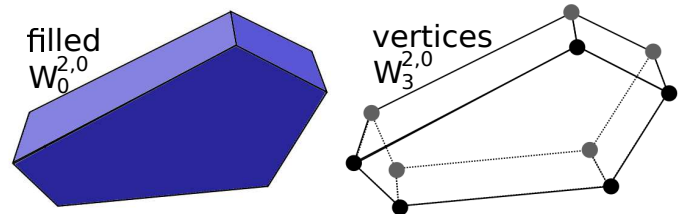}
    \caption{Visualization of two of the six Minkowski tensors.}
    \label{fig:minkvalexpl}
\end{figure}

We will now examine two examples of Minkowski tensors closer. We start with a Voronoi cell as shown
on the left side of figure \ref{fig:minkvalexpl}. It can be described as a convex body $K$ in Euclidean 3D space 
with a bounding surface $S$. The Minkowski Tensor $W_0^{2,0}$ is defined as:
\begin{equation}
  \label{eq:w020}
  \mathbf{W}_0^{2,0} \,=\, \int_K \vec{r}^2 \diff V
\end{equation}
where $\vec{r}$ is a position vector and the integral is over the volume of the filled cell.
$W_0^{2,0}$ can therefore be compared to the moment of inertia tensor.

The right side of figure \ref{fig:minkvalexpl} illustrates $W_3^{2,0}$, which is defined as:
\begin{equation}
  \label{eq:w320}
  \mathbf{W}_3^{2,0} \,=\, \int_S \vec{r}^2 G (\vec{r}) \diff A
\end{equation}
$G(\vec{r})$ is the Gaussian curvature of the surface, which is only non-zero on the 
vertices of the Voronoi cell. $W_3^{2,0}$ is akin  a moment of inertia tensor  
where only the vertices of the Voronoi cell contribute.

\begin{figure}[t]
\begin{center}
    \includegraphics[width=0.5\textwidth]{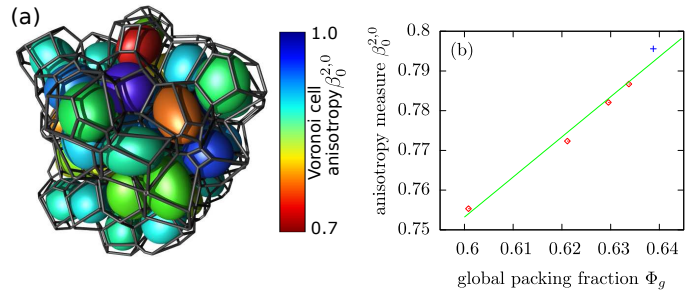}
\end{center}
\caption{Quantifying the shape anisotropy of Voronoi cells. 
   (a) The same subset of the Voronoi diagram shown in figure \ref{fig:voronoi_3d}a. 
    The beads have been replaced by ellipsoids that match the anisotropy of the Voronoi cells. 
    The anisotropy is quantified by $\beta_0^{2, 0}$ and is also indicated by the color of the ellipsoids.
    (b) Average anisotropy measure $\beta_0^{2, 0}$ as a function of the global packing fraction $\phi_g$ for systems of monodisperse spheres. 
   The example tomogram that is provided with this document is marked as a blue cross.}
   \label{fig:voronoi_shape}
\end{figure}

Once the Minkowski tensors have been computed for a Voronoi cell, its shape anisotropy can be quantified
as the ratio $\beta_\nu^{rs}$ between the smallest to the largest eigenvalue of the tensor $W_\nu^{r,s}$.
Thus, a cell with $\beta_\nu^{rs} = 1$ is isotropic while smaller values indicate larger anisotropies.

Figure \ref{fig:voronoi_shape}a shows the same Voronoi diagram as figure \ref{fig:voronoi_3d}a but the spheres have been 
replaced by ellipsoids with an aspect ratio identical to the $\beta_0^{2,0}$ values of the Voronoi cells.
Figure \ref{fig:voronoi_shape}b demonstrates that for packings of monodisperse spheres  the mean eigenvalue 
ratio $\beta_0^{2,0}$ increases linearly $\phi_g$. For crystalline packings  $\beta_0^{2,0}$ reaches unity (corresponding to isotropic
Voronoi cells). In the opposite limit of an equilibrium fluid with vanishing density, $\beta_0^{2,0}$ approaches 
the value 0.37, which is characteristic for Poisson point processes.\cite{kapfer2010local}
It has also been shown that the anisotropy of the Voronoi cells is 
independent of preparation or gravity.\cite{schroeder-turk:10,kapfer2012jammed}

\subsubsection{Computing Minkowski tensors}

The eigenvalues of the Minkowski tensors can be calculated using the program \texttt{karambola} 
\cite{turk2013karambola, mickel2008robust, schroder2010tensorial} using the files written by \texttt{pomelo} 
as input. Details can be found in appendix \ref{sec:appkarambola}.

\texttt{karambola} will also write a file containing the scalar Minkowski value $W_0^{0,0}$ which corresponds to the volume of the Voronoi cell
(in units of voxels$^3$). The average sphere diameter has already been determined from the radial distribution function 
in section \ref{sec:pair_correlation}, allowing us to compute a precise particle volume, again in units of voxels$^3$
Thus the output of \texttt{karambola} can also be used to calculate the local packing fraction of all particles 
using equation \ref{eq:phi_l}.

\section{Outlook}
Real world granular materials rarely consist of monodisperse spheres. Their
particles are sometimes concave, almost always polydisperse, 
and often only approximately monoschematic; coffee beans are a good 
example of this combination. There are two technological developments 
that will help granular physics to close this gap: 3D printing and advances 
in image processing. 

3D printing has been around for a couple of decades in academic and industrial settings. 
But in the last years a number of innovative startups have entered the market and 
considerably lowered the price tag on these machines. One way of moving away from 
monodisperse spheres is to print particles with complex or even concave shapes (such as
jacks or dolosse) and of varying size.\cite{athanassiadis:13,scholz:16}
The precision of the 3D printing process will then allow to still make use of the knowledge of their shape
during image processing.  

Alternatively, we can directly use natural materials such as sand and improve our 3D image processing to obtain
trustworthy segmentation results. The development of level-set methods in the group Gioacchino  Viggiani
marks an important step in this direction.\cite{vlahinic:14,kawamoto:16,vlahinic:16} 
There exist also hitherto unrealized synergies with the community 
studying multiphase flow in porous media. 
In their x-ray tomography images of fluid distributions in porous media  they are
confronted with very similar segmentation problems.\cite{wildenschild:13,leu:14,andrew:15}

Taking a step back we note that the spread of X-ray tomography in granular physics 
is part of a larger shift in physics and science in general. 
Modern measurement devices allow the acquisition of larger and larger data sets. 
Which in turn enables us to test ever complexer hypothesis. However, the 
amount of data processing and reduction necessary to do so increases almost at the same 
rate as the amount of data.  This article is a good example. Starting from a 300 MByte
large volume file, we identify the roughly 150  kByte large list of coordinates to end 
up finally with several tens of Bytes telling us the volume fraction, contact number, and 
Voronoi cell anisotropy of the sample.

This massive increase of the importance of data processing challenges the historically grown
model of scientific publishing because it strongly increases the danger of propagating undetectable errors 
made during this data processing step. Such errors might be innocent mistakes made by faithful 
scientists. Or they could be intentional cherry-picking to increase the impact factor of the result.  

We believe that we need to change our understanding how scientific publishing as a process works:
Raw data {\it and} the software needed to process them have to become an integral part of any 
scientific publications. 
There is already an established infrastructure for publishing the raw data in the form 
of open data repositories such as \texttt{Dryad}\cite{url_dryad} and \texttt{Zenodo}\cite{url_zenodo}.
To open up the analysis, it is a good first step to publish the source code under an open source license 
such as the GPL. However, we will also need new tools and protocols that facilitate to document
and preserve the individual choices (e.g. parameter values for image processing) 
made during the analysis. But these are technological, solvable problems. Getting the scientific culture
to open up to this level of sharing will likely be more difficult, but at the same time it seems 
unavoidable. The intrinsic strength of this vision is best captured in a recent comment by Tal Yarkoni:\cite{yarkoni:16} 
I hope in 20 years we’ll be amazed that scientists once blindly trusted results they couldn’t press the “run” button on.

\section{Supplementary Material}
The supplementary material consists of source code, data and binaries and will be discussed in detail in the appendix \ref{sec:app_analyzing}.

\section*{Acknowledgments}
We would like to thank Philipp Sch\"onh\"ofer, Manuel, Baur, Max Neudecker, Fabian Schaller, 
Gerd Schr\"oder-Turk, and Song-Chuan Zhao for helpful discussions. This article started as a talk  
given at the Spring School ``Imaging Particles'' held in Erlangen, Germany in April 2016 and
funded by German Science Foundation (DFG) through the Cluster of Excellence "Engineering of Advanced Materials" EXC-315.
The development of the software described in this article was funded by the German Research Foundation (DFG) Forschergruppe FOR1548 "Geometry and Physics of Spatial Random Systems" (GPSRS).
Figure \ref{fig:pair_correlation} (a) has been reproduced from Ref. \cite{baranau:16} with permission from the Royal Society of Chemistry. 
\bibliography{simon,matthias,x-ray,rsi}

\appendix

\section{Analyzing the example volume dataset}
\label{sec:app_analyzing}
This appendix lists in a concise way all the steps necessary to analyze the example data set 
accompanying this paper.

\subsection{Obtaining code and data}

\begin{enumerate}
    \item Download the example volume from \cite{demo_url} and store it as \texttt{volume.raw} in your working directory. 
       This volume file has been acquired with a Nanotom Tomograph from {\it GE Sensing and Inspection}. 
       Imaging conditions were 120 kV, 120 $\mu$A with 6 averages, each with an exposure time of 250 ms and a total of 1400 projections. 
       The original volume had a size of 1132$\times$1132$\times$1152 voxels. After binning and cropping, the volume was reduced to a total 
        of 512$\times$512$\times$512 voxels.
        The sample consists of spheres of POM with an diameter of 3.5 mm that have been tapped 16000 times with a maximum acceleration of 
        $\Gamma=2g$ at a frequency of 3 Hz. 
         
    \item For viewing the volume file, download and install \texttt{Fiji} from the official web site.\cite{url_fiji}

    \item The program \texttt{volume2positionList} is needed to extract a list of particle positions from the tomogram. 
      The preferred way to obtain it is to download the source code from \cite{url_git_v2p} and store in a sub folder \texttt{code} 
      in your working folder.  To compile the program, change to the \texttt{code} directory and call \texttt{make} there.

    \item While compiling \texttt{volume2positionList} from the source code should be considered the best option, we do
      understand that this might create an additional barrier for some users. Therefore we provide as an alternative 
      a static linked binary with the supplemental material. It should run out of the box on most 64-Bit Linux PC's. 
      The supplemental material contains also static linked binaries for the other three analysis programs discussed below.

    \item To calculate the pair correlation function and the global contact number, you will need to download the source code 
     for the program \texttt{raps} from \cite{url_git_raps} and store it into a folder called \texttt{raps} in the working directory. 
     \texttt{raps} can be compiled with the included \texttt{build.bsh} shell script. For further instructions please refer to the 
     included \texttt{README.md}. 
    
   \item Calculating the Set Voronoi diagrams, requires the program \texttt{pomelo}.
     The corresponding source code can be downloaded from \cite{url_git_pomelo} and then stored in folder called 
     \texttt{pomelo} within the working directory. In order to compile \texttt{pomelo}  
       with the included makefile, a compiler that supports most C++11 features is required, e.g. \texttt{clang++ 3.5.0-10}.  
       Further information about compiling \texttt{pomelo} and the included examples are given in the file \texttt{README.md}.

    \item If you want to characterize the Voronoi volumes using Minkowski tensors, you will need to download 
\texttt{karambola} from \cite{url_karambola}. After storing it in a folder called \texttt{karambola}, you can 
compile it with the included makefile. 
\end{enumerate}

\subsection{Visualization using \texttt{Fiji}}
\label{sec:appfiji}

\texttt{Fiji} is an image processing package that is built on top of \texttt{ImageJ},\cite{url_imagej} an open source image processing program.
\texttt{Fiji} is run in an java virtual machine.
On some systems, the virtual machine starts with a limited amount of memory which is not enough for most 3D image 
applications.\footnote{To find out the amount of memory available to \texttt{Fiji}, check the settings in 
\textit{Edit} $\rightarrow$ \textit{Options} $\rightarrow$ \textit{Memory and Threads}.}
To tell the machine to use more memory a command line command has to be typed prior to starting \texttt{Fiji}. 
The following command will increase the memory limit to 2000MB, which is sufficient for the applications explained in this document.

\verb%export _JAVA_OPTIONS="-Xmx2000M -Xms2000M"%

\noindent After that, start \texttt{Fiji} from the same command line.  

To open a tomography file, you have to specify a number of parameters in the \textit{open raw} dialogue. The parameters for the sample tomogram are given in figure \ref{fig:fiji} b).
These parameters can be divided into three different groups: volume specific, visualisation choices and machine specific. The values for the sample data are given in square brackets.

The volume specific parameters are\textit{Width}, \textit{Height} and \textit{Number of Images}.
These parameters depend on both the settings which the tomography was taken with (size in $x$, $y$ and $z$) and the choices
made during the reconstruction of the tomogram.  
This information is typically contained in a text file saved together with the volume file. Or it can be obtained from
the export dialog for the volume in the reconstruction software. [use 512 for each dimension]

\noindent Machine specific parameters will be identical for all volumes exported from the same 
reconstruction software and should be documented in the manual belonging to that software:
\begin{description}
    \item[Image Type] The data type of the individual voxels are stored in. [16-Bit Unsigned]
    \item[Offset] Size of a possible header preceding the volume data (in bit). Use 0 for none. [0]
    \item[Gap]  Size of a possible header preceding each slice (in bit). Use 0 for none. [0]
    \item[Byte Order]Integers can be stored in two different ways. This switch reverses the order of the binary values of each datatype. 
      [Check] 
    \item[Open All Files] Some software saves the volume not as one single file but as series 
       of multiple files (e.g. the different slices of a tomogram). This option will load all of them. 
       [Uncheck]
\end{description}

\noindent Finally, there are two choices which influence how the volume is visualized:
\begin{description}
    \item[White is zero] In x-ray tomographies the value of a voxel is proportional to the local absorption coefficient: 
white corresponds to high absorption. This switch would invert this relationship. [Uncheck] 
    \item [Use virtual Stack] 3D volume dataset can become several GByte large.
    If a PC has not enough RAM to load the complete image data instantly to the memory, 
    this switch allows to just load the slice which is presently 
     accessed for viewing or processing.  This option has a serious impact on performance. 
     [Uncheck]
\end{description}

\subsection{Particle detection on the example data}
\label{sec:apppartdet}
The program \texttt{volume2positionList} identifies all spheres in the volume data set 
(using the methods described in section \ref{sec:find_particles})
and then  exports their particle positions as a text file. 
The parameters of \texttt{volume2positionList} are \texttt{[inFileName] [outFolderName] [nx] [ny] [nz] [erosion depth] [sigma\_g] [sigma\_p]}.

The parameters \texttt{inFileName} and \texttt{outFolderName} must be adapted to your filesystem structure. 
\texttt{nx}, \texttt{ny}, \texttt{nz} are the size of the volume file (512 for the example). 
\texttt{erosion depth} refers to the erosion step described in section \ref{sec:centroidMethod} 
and should be set to a value of 5 voxels. 
The \texttt{sigma} parameters correspond to the respective values in the bilateral filter step $\sigma_g$ and $\sigma_p$.

All output will be written to the output folder \texttt{outFolderName}. 
The program will also create a subdirectory \texttt{images} which contains slices at three different heights through the volume at different processing steps. 
The list of particle positions will be written to a file called \texttt{particles\_centroids\_ed5.dat}.
To run the program on the example data, use 
\begin{center}
    \texttt{volume2positionList volume.raw output 512 512 512 5 1.75 2000}
\end{center}

After running \texttt{volume2positionList}, the first two lines of \texttt{particles\_centroids\_ed5.dat}
in the the analysis directory should look like this:

\noindent \verb%255.106 255.061 255.083 0   90197774%
          \verb%345.993 190.616 38.1658 472 10746 %

The first line represents the air cluster, the second line gives the coordinates (column number 1 to 3) 
of the first detected sphere.  Column 4 lists the id  of the particles, column 5 is the volume of the particle.

\subsection{Pair correlation calculation with \texttt{raps}}
\label{sec:appraps}
The program \texttt{raps} receives its parameters by loading a \texttt{data.set} file.
For our example this file has to contain the following line:\\
\texttt{particles\_centroids\_ed5.dat ip\_example 0.005 27.17}. 

The first entry is the filename of the position list, the second entry an identifier that will be used to label the output files. 
The number in the third column is the bin size $\Delta r$  (cf.~equation   \ref{ref:eq_gofr}) used for computing the pair correlation function.
The last number is the particle diameter, which is needed to compute the raw data used in 
the CNS function fit described in the next section. The value of 27.17 is the correct 
particle diameter for the demo data; for other raw data you should provide an educated guess 
which can then be updated in a second iteration once you have measured the particle diameter as 
described in the next section. 

After creating the the configuration file \texttt{data.set} you can call the program (from the analysis folder) with the command 
\begin{center}
    \texttt{../raps/RAPS data.set}. 
\end{center}

\texttt{raps} will calculate structural properties of the packing, 
such as the numerator in equation \eqref{ref:eq_gofr}, the pair correlation function $g(r)$ (\texttt{pc\_ip-example.dat}) 
(the normalisation has to be performed by the user)
and the raw data required to compute the contact number (\texttt{cns\_ip\_example.dat}).
Each file will contain information how to plot it using \texttt{gnuplot}.\cite{url_gnuplot}

Computing the mean diameter of the particles is the first step in the next section and is described there.

\subsection{Determining global contact numbers}
\label{sec:appcns}
Determining the global contact number using the ensemble based method, as described in section \ref{sec:cns},
is a two step process. While you can use any program capable of fitting equations 
to data to perform this task, we provide for convenience a gnuplot \cite{url_gnuplot} skript named \texttt{find\_contacts.plt}
in the supplemental materials. Please note that fit-programs such as gnuplot normally use a standard Levenberg-Marquardt algorithm.
Which means that the starting values for the fit parameters have to be chosen carefully, as the algorithm can easily get stuck in a erroneous, local minimum.

The first step is to compute the mean particle diameter by fitting the first peak of the pair correlation function 
with a Gaussian distribution. You will find the raw data of the pair correlation function in the file  
\texttt{pc\_ip-example.dat} written by \texttt{raps}. The width of the fitting window should be limited to include only the 
upper half of the first peak of $g(r)$.  Fitting the $g(r)$ values of our demo dataset provides a mean sphere 
diameter $d_\textnormal{mean}$ of 27.177 voxels (cf. figure \ref{fig:pair_correlation}a).  

The second step is to fit the contact number scaling function in equation \ref{eq:cnsfunction} to the data in the file
\texttt{cns\_ip\_example.dat}. First you should only determine the width of the error function $\sigma$ by fitting $C(x)$ to the left part of the data, i.e.~all points smaller than   $d_\textnormal{mean}$. 
This requires a reasonable estimate for $Z$.
Our fit yields  $\sigma$ = 0.0569. Then the full CNS function can be fitted to determine the contact number $Z$ 
(and the pre factor of the linear offset). The upper range of this fit should be chosen as the real particle diameter 
plus 5 times the width of the error function.  The contact number of our demo dataset is 5.4.

\subsection{Set Voronoi calculation on sphere packings}
\label{sec:apppomelo}
\texttt{pomelo} is a open source program, licensed under \texttt{GPLv3} that uses the \texttt{voro++} 
library (licensed under a modified \texttt{BSD} license).\cite{url_voro++, rycroft2009voro++} 
\texttt{pomelo} can be used to calculate the Set Voronoi cells from a list of sphere centers. 
The parameters are:
\texttt{pomelo -mode SPHERE -i [inFileName] -o [outFolderName]}.
In this command, \texttt{-i inFileName} should be the path to the \texttt{particles\_centroids\_ed5.dat} file created by \texttt{volume2positionList} program and \texttt{-o outFolderName} should be the path to the output folder.
Note that for monodisperse spheres (\texttt{-mode SPHERE}) the Voronoi and the Set Voronoi diagram are identical. 
Thus it is sufficient to use the centers of each sphere to create the Voronoi Diagram.
The command (out of the analysis folder) to run pomelo on the example data is
\begin{center}
    \texttt{../pomelo/bin/pomelo -mode SPHERE -i particles\_centroids\_ed5.dat -o pomeloOut}
\end{center}

\texttt{pomelo} creates by default five files in the output folder. 
The file \texttt{reduced.xyz} contains the vertices, faces and cells of the Set Voronoi cells, the file
\texttt{surface\_triangulation.xyz} contains the surface of the particle, or in case of spheres, where 
the Set Voronoi cells coincide with the normal Voronoi cells, the centers of the particles. 
Both files can be plotted with \texttt{gnuplot} using 

\noindent \texttt{splot "surface\_triangulation.xyz" u 2:3:4 w p} 

or

\noindent \texttt{splot "reduced.xyz" u 2:3:4 w lp}

The files \texttt{cell.poly} and \texttt{cell.off} contain the same information as the file  
\texttt{surface\_triangulation.xyz} but are intended to be used with  
\texttt{karambola} (\texttt{cell.poly}) and \texttt{geomview} (\texttt{cell.off}).

Finally, the file \texttt{setVoronoiVolumes.dat} contains the volume of each (Set) Voronoi cell. 
This information can be used to calculate the local packing fraction.

\subsection{Calculate the Minkowski Tensors of the Set Voronoi cells}
\label{sec:appkarambola}
The Minkowski tensors and their Eigenvalues of the Set Voronoi cells can be calculated by the program \texttt{karambola}
using the output of \texttt{pomelo} (see section \ref{sec:pomelo}) as input file. 
The necessary command (from the \texttt{pomeloOut} folder) is:
\begin{center}
    \texttt{../../karambola/karambola -i cell.poly -o karambolaOut --labels --reference\_centroid}
\end{center}

The parameters of \texttt{karambola} are the path to the poly file created by pomelo
(\texttt{-i [input poly file]}), the output folder (\texttt{-o [output folder]}), and the
the switch \texttt{--labels } which tells karambola to use the labels of the individual cells. 
Otherwise all cells would be combined into one large cell.
The \texttt{--reference\_centroid} option tells karambola to use the centroid of the cell as the reference point, in contrast to the default behavior where the origin of the sample will be used as the reference for calculating the Minkowski Tensors.

Karambola's output consists of multiple files. 
The file \texttt{w000\_w100\_w200\_w300} for example contains the information about the 
volume $W_0^{0,0}$, the surface area $W_1^{0,0}$, integral mean curvature $W_2^{0,0}$ and the integral Gaussian curvature (genus) $W_3^{0,0}$ of each Set Voronoi cell.
With the particle volume $W_0^{0,0}$, the local packing fraction of each particle can be calculated.
The file \texttt{w020\_eigsys} contains the Eigenvalues and corresponding Eigenvectors of the $W_0^{2,0}$ Minkowski Tensor for each of the cells.


\section{The Hoshen Kopelman algorithm}
\label{sec:hk}
\begin{figure}[ht]
    \centering
    \includegraphics[width=0.5\textwidth]{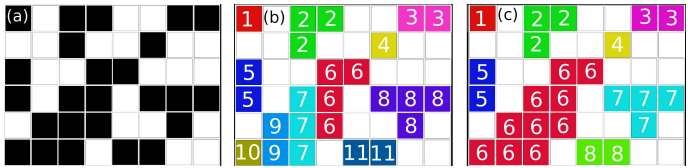}
    \caption{Applying the Hoshen Kopelman algorithm on a two-dimensional binary image. 
       (a) binarized image. 
       (b) first step of the algorithm, before resolving the indirections. 
       (c) final, labeled image.
     }
    \label{fig:hk}
\end{figure}
Clusters of pixels or voxels can be identified with the \textsc{Hoshen Kopelman} algorithm.\cite{hoshen1976percolation}
Figure \ref{fig:hk} explains it on a two-dimensional image where we assume that the clusters are composed of black pixels. 
The algorithm iterates through all pixels of the binary image 
(figure \ref{fig:hk}(a)) starting at the top left corner and processing line by line from left to right. 
If the pixel under consideration is a white pixel, the algorithm proceeds to the next pixel.
When a black pixel is encountered, it is assigned a cluster id which is determined by considering the left and top neighbor of that pixel,
which have both already been visited by the algorithm.

If both are white pixels or pixels outside the image boundary, a new cluster id is assigned to the pixel under consideration.
This is e.g.~the case for pixel with cluster id 4 in figure \ref{fig:hk}(b), respectively the top pixel of cluster number 5.
If either the left or the top neighbor is black, the cluster id of this pixel is used. 
An example is the first black pixel in row 2 which is assigned the cluster id of 2 stemming from the pixel above it.
If both left and top pixel are black, the lower of the cluster IDs of the neighbors is used as it is the case for the 
second black pixel in the lowest row of figure \ref{fig:hk}(b). 
All touching Clusters with different IDs are detected in this step and connectivity is stored.

Then in a second step, connected clusters are merged using the connectivity information.
In this case, the clusters with IDs 6, 7, 9, 10 are all assigned value 6. 
The final labeled image is shown in figure \ref{fig:hk}(c).

\end{document}